\pdfoutput=1

\documentclass[11pt]{article}

\usepackage{ACL2023}
\usepackage{tcolorbox}
\usepackage{amsmath}
\usepackage{colortbl}

\usepackage[capitalize,noabbrev,nameinlink]{cleveref}
\usepackage{mdwlist}

\newcommand{\method}{\textsc{MGR-LF++}\xspace}

\definecolor{Gray}{gray}{0.85}

\newenvironment{example}
    {\begin{tcolorbox}[colback=gray!5,colframe=gray!50!black,title=Example]}
    {\end{tcolorbox}}

\newtheorem{definition}{Problem}

\usepackage{times}
\usepackage{latexsym}
\usepackage{pifont}
\usepackage{newunicodechar}
\usepackage{algorithm}
\usepackage{multirow}
\usepackage{array}
\usepackage{subcaption}
\usepackage{bbm}
\usepackage{enumitem}
\usepackage{booktabs}     
\usepackage[noend]{algpseudocode}

\usepackage[T1]{fontenc}

\usepackage[utf8]{inputenc}

\usepackage{microtype}
\usepackage{amsmath}
\usepackage{amssymb}
\usepackage{graphicx}
\usepackage{xspace}

\usepackage{inconsolata}

\newunicodechar{✓}{\ding{51}}
\newunicodechar{✗}{\ding{55}}

\usepackage{ulem}

\title{Beyond Unimodal Boundaries: Generative Recommendation with Multimodal Semantics}

\author{ Jing Zhu$^{\ 1}$\quad Mingxuan Ju$^{\ 2}$ \quad Yozen Liu$^{\ 2}$ \quad Danai Koutra$^{\ 1}$\\ \quad \textbf{Neil Shah}$^{\ 2}$ \quad \textbf{Tong Zhao}$^{\ 2}$\\
          $^{1}$ University of Michigan, Ann Arbor \\ 
              $^{2}$ Snap \\ 
         \texttt{\{jingzhuu, dkoutra\}@umich.edu} \\
        \texttt{\{mju, yliu2, nshah, tzhao\}@snapchat.com}}

\begin{document}
\maketitle
\begin{abstract}

\begin{figure*}[t!]
\centering

  \includegraphics[width=0.8\textwidth]{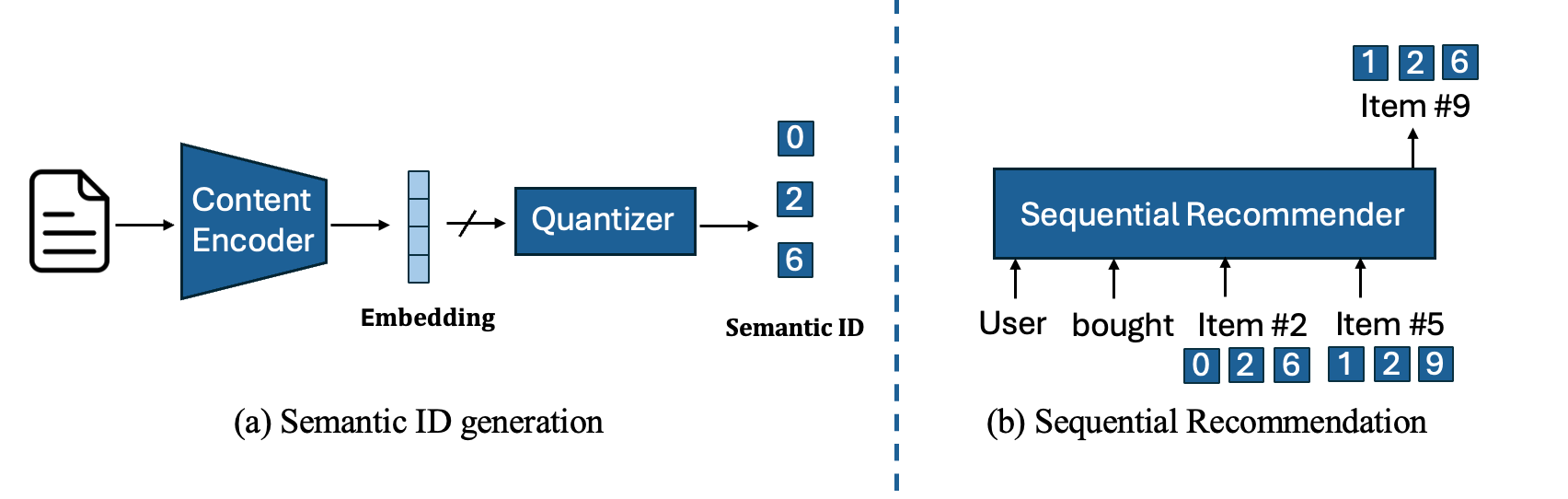}

\caption{\textbf{Overview of a Typical Generative Recommendation Pipeline.} (a) Each item's semantic ID is generated as a list of discrete IDs based on its content. (b) These generated semantic IDs are then used to represent items within the sequential recommender.}
 \label{fig:overview}
 
\end{figure*}

Generative recommendation (GR) has become a powerful paradigm in recommendation systems that implicitly links modality and semantics to item representation, in contrast to previous methods that relied on non-semantic item identifiers in autoregressive models. 
However, previous research has predominantly treated modalities in isolation, typically assuming item content is unimodal (usually text). We argue that this is a significant limitation given the rich, multimodal nature of real-world data and the potential sensitivity of GR models to modality choices and usage. 
Our work aims to explore the critical problem of \textit{Multimodal Generative Recommendation (MGR)}, highlighting the importance of modality choices in GR frameworks. 
We reveal that GR models are particularly sensitive to different modalities and examine the challenges in achieving effective GR when multiple modalities are available.
By evaluating design strategies for effectively leveraging multiple modalities, we identify key challenges and introduce \method, an enhanced late fusion framework that employs contrastive modality alignment and special tokens to denote different modalities, achieving a performance improvement of over 20\% compared to single-modality alternatives.

\end{abstract}

\section{Introduction}

Recommender systems help users discover content of interest and  play an essential role in various domains such as videos~\cite{covington2016deep}, e-commerce~\cite{schafer2001commerce}, and products~\cite{geng2022recommendation}. Recently, the \textit{generative recommendation (GR)}  which directly generates a ranked list of identifiers (IDs) for a given query using autoregressive models (e.g., large language models) has become increasingly popular~\cite{sun2024learning, rajput2023recommender, zheng2024adapting, deldjoo2024review}. As shown in \cref{fig:overview}, at the core of GR models is the construction of \textit{semantic IDs}, a sequence
of discrete IDs which capture the semantic meaning
of a document or item.
Compared to existing approaches in sequential recommendation that typically assign a unique ID to each item, or traditional non-sequential recommendation approaches which often use random or frequency-based strategies for assigning IDs to items,
semantic IDs aim to encapsulate the underlying characteristics or available modality information from items as an inductive prior to enable efficient retrieval and generalization to unseen items~\cite{rajput2023recommender, lin2024bridging, wang2024learnable, zheng2024adapting}. 

Despite the advancement of GR methods and prevalence of multimodal contents in recommendation systems~\cite{liu2024multimodal},
prior works have largely treated modality information exclusively, by inferring semantic IDs from unimodal (and often textual) content. However, GR is strongly grounded in the quality and availability of modality information, and treating it as unimodal raises questions about its effectiveness and sensitivity in multimodal environments, which are prevalent in real-world scenarios. Motivated by this limitation, we focus on the following fundamental question: %

\begin{center}
    \textbf{\textit{What are the key principles driving effective\\ multimodal generative recommendation (MGR)?}}
\end{center}

\begin{figure*}[t!]
\centering

  \includegraphics[width=0.8\textwidth]{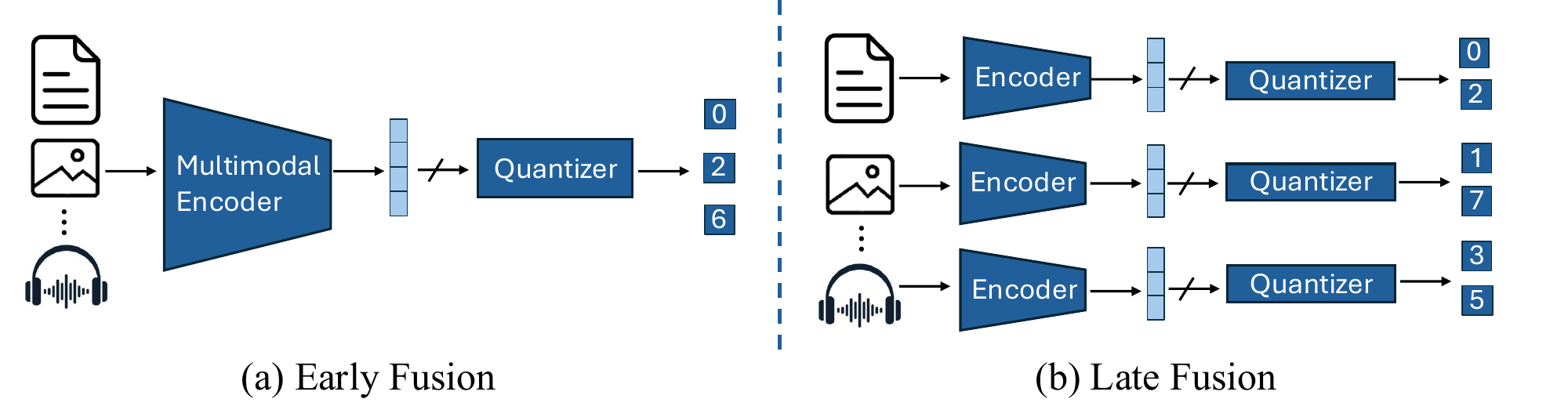}

\caption{\textbf{Naive strategies for extending unimodal generative recommendation to multimodal scenarios.} Early fusion (MGR-EF) generates a unified list of semantic IDs capturing the semantics across all modalities. Late fusion (MGR-LF) generates separate semantic IDs for each modality and combines them after generation.}

  \label{fig:fusion}
  \vspace{-0.1in}

\end{figure*}

To answer the question, we first explore the effectiveness of existing generative recommendation methods under multimodal environments. To enable their capability of taking multimodal data, we leverage early fusion and late fusion strategies (as illustrated in \cref{fig:fusion}).
Early fusion employs a single multimodal encoder, such as ImageBind~\cite{girdhar2023imagebind}, to encode content from multiple modalities simultaneously, generating a unified set of semantic IDs that capture the semantics across all modalities. However, in practice, we found that one modality tends to dominate and there is significant information loss.  In contrast, late fusion generates separate semantic IDs for each modality and combines them after the ID generation, but our study highlights that it lacks a way to match corresponding semantic IDs across modalities.

Motivated by these observations, we first validate the importance of preserving the distinct information from each modality, which requires maintaining separate IDs, as done in late fusion. Moreover, to increase the effectiveness of the late fusion scheme, we propose two designs that alleviate the modality correspondence issue:
(1) contrastive modality alignment training that directly matches corresponding semantic IDs across different modalities, and (2) special tokens that separate sequential IDs from different modalities.

Our contributions are summarized as follows: 
 
\begin{itemize}[leftmargin=*]
    \item \textbf{Problem Formulation:} We present the first systematic study of multimodal generative recommendation or MGR, a practical but understudied research problem. We empirically evaluate early fusion and late fusion solutions, and identify the challenges that prevent existing GR solutions from working well under multimodal scenarios.

    \item \textbf{New Late Fusion Framework:}
    Based on our findings, we introduce \method, an enhanced late fusion framework with two key designs that allow for maintaining the different modalities and aligning them for best performance.

    \item \textbf{Empirical Analysis:} Through extensive experiments with 6 baselines on 3 datasets, we show that carefully incorporating multimodal information into GR can significantly improve the performance (up to 20\%) compared with its unimodal counterparts.

\end{itemize}

\section{Preliminaries}

We first present a typical generative recommendation pipeline, as illustrated in \cref{fig:overview}. This pipeline comprises two distinct stages.

\paragraph{Stage 1: Semantic ID Generation.} This stage involves encoding item content features using an off-the-shelf encoder to produce embedding vectors. These vectors are then quantized into a tuple of semantic codewords using an auto-encoder framework (e.g., RQ-VAE). The resulting tuple of codewords is referred to as the item's semantic ID.

\paragraph{Stage 2: Sequential Recommendation.} Given a user's interaction history in the form of a sequence of semantic IDs (item$_1$, ..., item$_n$), the recommender system's task is to predict the next item, item${_{n+1}}$, also in the form of a semantic ID.

\section{The MGR Problem}
\label{sec:studi}

In this section, we begin by presenting the multimodal generative recommendation (MGR) problem, followed by a discussion of the unique challenges it poses and our proposed solutions to address these challenges.

\begin{definition}[MGR]
Under a sequential recommendation problem setting, given all items with their associated multi-modal contents, how can we:
\begin{itemize*}
    \item Generate a list of discrete semantic IDs $(c_{i_0}, ..., c_{i_{m-1}})$ that effectively capture the semantics of multimodal contents, one for each item $i$?
    \item Design a sequential recommendation model that  effectively integrates these semantic IDs for improved performance?
\end{itemize*}
 \end{definition}

\begin{table}[!t]
\centering
\caption{\label{table:ami} \textbf{MGR-EF struggles to preserve semantics from all modalities, resulting in information loss.}
We present Adjusted Mutual Information (AMI) scores between modalities.
The AMI scores between text and multimodal consistently outweigh others across datasets,
indicating that textual information dominates in the multimodal representations, while other modalities' contributions are diminished.
}
{\small
\begin{tabular}{lccc}
\toprule
\textbf{Datasets} & \textbf{Img vs.} & \textbf{Img vs.} & \textbf{Txt vs.} \\ 
& \textbf{Txt} & \textbf{Multimodal} & \textbf{Multimodal} \\ \midrule
\textbf{Toys} & 0.2765 & 0.3892 & \colorbox{Gray}{0.4829} \\
\textbf{Beauty} & 0.2920 & 0.3937 & \colorbox{Gray}{0.5539} \\
\textbf{Sports} & 0.3976 & 0.5105 & \colorbox{Gray}{0.6351} \\
\bottomrule
\end{tabular}
}
\end{table}

To extend generative recommendation pipelines into multimodal environments, we begin by exploring two fundamental strategies: early fusion and late fusion. These approaches aim to effectively integrate multimodal content for generating semantic IDs, as illustrated in \cref{fig:fusion}.

\subsection{MGR-EF: Naive Early Fusion Framework}

As shown in \cref{fig:fusion}(a), the early fusion strategy employs a single multimodal encoder to process content from multiple modalities concurrently. The resulting unified representation is then used to generate a single set of semantic IDs that encapsulate the semantics across all modalities. This approach ensures that inter-modal interactions are captured during the encoding phase, potentially leading to richer representations. However, early fusion may face challenges such as over-reliance on dominant modalities.

\begin{table}[!t]
\centering
\caption{\label{table:predict_overlap} \textbf{MGR-EF produces similar prediction results to its unimodal counterparts, indicating potential information loss.}
We present the percentage of overlapping predictions across modalities, revealing the frequency with which examples yield identical prediction results across different modality inputs. 
}
{\small
\begin{tabular}{lccc}
\toprule
\textbf{Datasets} & \textbf{Img vs.} & \textbf{Img vs.} & \textbf{Txt vs.} \\ 
& \textbf{Txt} & \textbf{Multimodal} & \textbf{Multimodal} \\ \midrule
\textbf{Toys} & 0.3287 & 0.3881 & \colorbox{Gray}{0.4612} \\
\textbf{Beauty} & 0.4724 & \colorbox{Gray}{0.4911} & 0.4866 \\
\textbf{Sports} & 0.2775 & 0.3675 & \colorbox{Gray}{0.4584} \\
\bottomrule
\end{tabular}
}
\vspace{-0.3cm}
\end{table}

\begin{figure*}[t!]
\centering
\includegraphics[width=0.9\textwidth]{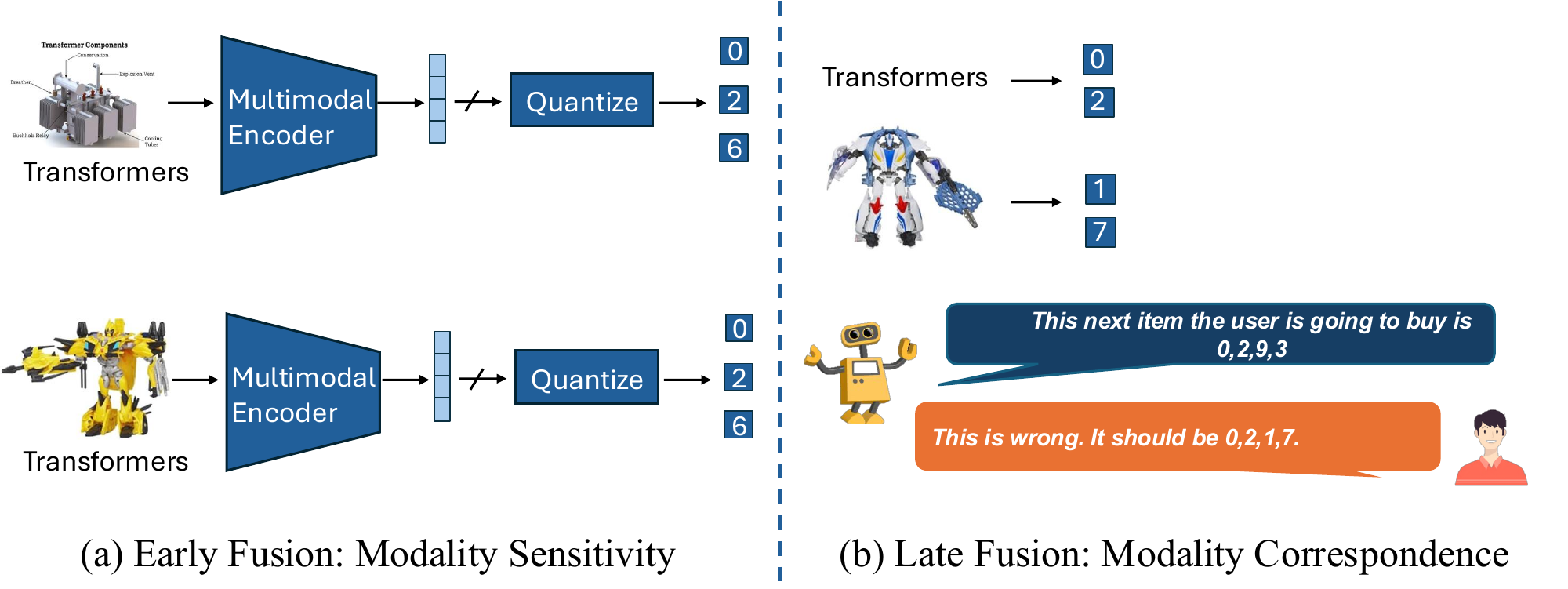}
\caption{\textbf{Key challenges in the MGR problem.} (a) Modality Sensitivity: Despite both items being transformers, they are distinct products. The semantic ID generated through early fusion fails to capture visual differences, resulting in information loss. (b) Modality Correspondence: The item's textual semantic ID is [0,2], with a corresponding visual semantic ID of [1,7]. While the sequential recommender correctly predicts the item as a transformer (by accurately predicting the textual semantic ID [0,2]), it fails to map this to the corresponding visual semantic ID [1,7].}
\label{fig:example}
\end{figure*}

\textbf{Finding 1: While the early fusion strategy is straightforward, it is susceptible to the modality sensitivity challenge, wherein the generated semantic IDs are dominated by one modality and fail to preserve the distinct semantics of multiple modalities.} As illustrated in \cref{fig:example}(a), one modality may dominate the encoded information, resulting in identical semantic IDs for items that differ in other modalities.

To quantify this dominance, we employ two metrics: the Adjusted Mutual Information (AMI) score (defined in \cref{sec:ami}) and the percentage of overlapping predictions in sequential recommendation, using text and image modalities as examples.

A higher AMI score indicates greater information overlap between generated semantic IDs, while a higher percentage of overlapping predictions suggests the sequential recommenders' inability to effectively utilize information from multiple modalities. As evidenced in \cref{table:ami,table:predict_overlap}, the comparison between text and multimodal inputs yields the highest AMI score and percentage of overlapping hits. This finding indicates that the MGR-EF approach is compromised by modality sensitivity challenges, making it suboptimal for effective generative recommendation.

To address this issue, we propose \textbf{generating separate semantic IDs for each modality} before combining them, as implemented in late fusion, to mitigate the modality sensitivity challenge. Formally, given a set of modalities $M={m_0, m_1, ..., m_k}$, we aim to generate a sequence of semantic IDs $C=[c_0, c_1, ..., c_n]$ where each $c_i$ corresponds to a specific modality $m_j$. This approach facilitates better preservation of modality-specific information and potentially more balanced representation across different modalities in the final recommendation process.

The development of a more balanced early fusion-based approach capable of generating multimodal semantic IDs without information loss remains an avenue for future research.

\begin{table}[t!]
\centering
\caption{\label{table:late_fusion} \textbf{MGR-LF fails correctly to generate the semantic IDs of multiple modalities sequentially.} We show Hits@5 of predicting semantic IDs of single modality versus multiple modalities.}
\resizebox{0.47\textwidth}{!}{
\begin{tabular}{lccc}
\toprule
\textbf{Datasets} & \textbf{Text Only} &\textbf{Image Only} & \textbf{Multimodal} \\ \midrule
\textbf{Toys} & \colorbox{Gray}{0.0598}   & 0.0495 & 0.0482\\
    \textbf{Beauty} & \colorbox{Gray}{0.0505} & 0.0417 & 0.0336 \\
\textbf{Sports} & \colorbox{Gray}{0.0195} & 0.0193 & 0.0184 \\
\bottomrule
\end{tabular}
}
\end{table}

\subsection{MGR-LF: Late Fusion Framework}

As illustrated in \cref{fig:fusion}(b), late fusion generates separate semantic IDs for each modality independently using unimodal encoders. These modality-specific semantic IDs are subsequently combined through simple aggregation techniques (e.g., concatenation). This approach allows each modality to contribute independently before integration.

Although generating separate semantic IDs for each modality preserves modality-specific information, it introduces a significant challenge: modality correspondence. As evidenced in \cref{table:late_fusion}, there exists a substantial performance degradation when predicting the entire multimodal semantic ID sequence compared to predicting only a subset (Text Only or Image Only) of the sequence.

We hypothesize that the root cause of these challenges stems from the following findings:

\paragraph{Finding 2: The sequential recommender struggles to map generated semantic IDs for one modality to their corresponding IDs in other modalities.} As depicted in \cref{fig:example}(b), while the sequential recommender correctly predicts the first two textual semantic IDs, it generates instead of the correct, indicating an inability to map the textual semantic ID sequence to its corresponding visual semantic ID sequence. Formally, when the sequential recommender generates a sequence of semantic IDs $C=[c_0, c_1, ..., c_n]$ for the next item, we observe that while it can correctly generate the first few IDs for one modality $c_{t_0}, .., c_{t_m}$, it fails to match the corresponding IDs $c_{v_0}, .., c_{v_m}$ in other modalities. This modality mismatch causes significant performance degradation, as the model would suffer from the false predictions of all modalities.

\paragraph{Finding 3: Multiple modalities introduce challenges in the conditional generation of semantic IDs.} Sequential recommenders typically generate semantic IDs autoregressively, with each subsequent ID based on previously generated ones. This approach is effective for unimodal data, where each new ID acts as a residual for the preceding ones. However, it becomes problematic in multimodal contexts, as the semantic information of one modality often does not depend on that of others.

Consider a sequence of semantic IDs $C=[c_0, c_1, ..., c_n]$. When $c_{i-1} \in m_f$ and $c_i \in m_j$ (i.e., consecutive semantic IDs belong to different modalities), the conditional generation of $c_i$ based on $c_{i-1}$ becomes challenging. This is because $c_i$ is not a residual for $c_{i-1}$, disrupting the conventional pattern of autoregressive generation.

\subsection{\method: Proposed Enhanced Late Fusion Framework}

Motivated by our findings from analyzing the naive MGR-EF and MGR-LF frameworks, we focus on late fusion that generates semantic IDs per modality (Finding 1) and propose an enhanced LF framework:\method that incorporates two key designs that address the challenges identified in Findings 2 and 3.

Specifically, to address Finding 2, we propose a new training paradigm called \textbf{Contrastive Modality Alignment}. We first fine-tune the sequential recommender by explicitly tasking it to identify corresponding IDs for other modalities. After the sequential recommender gains knowledge about matching text and visual IDs, we then fine-tune it on the sequential recommendation task. Below are examples of the training paradigm, using text and image as example modalities:

\begin{example}
\small
\textbf{Image-to-Text Prediction:} Source: Here we have an item with image id <img-1-40><img-2-80>.Please predict its corresponding text id.

Target: <txt-1-214><txt-2-250>.
\end{example}

\begin{example}
\small 
\textbf{Text-to-Image Prediction:} Source: Here we have an item with text id <txt-1-214><txt-2-250>.Please predict its corresponding image id.

Target: <img-1-40><img-2-80>.
\end{example}

Note that this contrastive alignment objective leverages only each item's semantic IDs, and no other information (e.g., user purchase history, item collaborative information) is used. Through this contrastive alignment, the model is able to match semantic IDs across different modalities.

To address Finding 3, we propose to incorporate \textbf{special tokens} $S={s_1, s_2, ..., s_{k-1}}$ into the semantic ID sequences. Each $s_i$ indicates a transition between modality $i-1$ and modality $i$. This results in a modified semantic ID sequence:
$$C=[c_0, .., c_{i}, s_1,  c_{i+1}, .., s_{k-1}, .., c_n].$$
By explicitly marking modality transitions, this approach facilitates smoother integration of multimodal information while maintaining the integrity of the autoregressive generation process.

\section{Experiments}
\label{sec:experiments}

In our empirical analysis, we aim to address the following research questions: 

\textbf{(RQ1)}~How effective is multimodal generative recommendation compared to unimodal generative recommendation?

\textbf{(RQ2)}~What is the impact of our \method on multimodal generative retrieval, compared with MGR-EF and MGR-LF? How effective is each design of \method?

\textbf{(RQ3)}~How does \method perform across various ID lengths and codebook sizes?

\subsection{Empirical Setup}

 \paragraph{Dataset} Due to the lack of publicly available multimodal data, our experiments primarily focus on two modalities: text and images. However, our analysis is generalizable to other modalities. 
Following ~\cite{jin2023lmindexer,rajput2023recommender}, we conduct experiments on three domains from Amazon review dataset \cite{ni2019justifying, hou2403bridging}, Toys,
Beauty, and Sports. For sequential recommendations, we keep the users and items that have at least 5 interactions in their history. %
We treat the last
interacted item by each user as the testing sample, the second to last
second item as the validation sample, and the
previous items as training samples. The text contents are the titles and descriptions of the
products and the visual contents are the high-resolution raw images of the products. Items without raw images are removed.
The statistics of the
datasets can be found in \cref{tab:datasets}. The data does not contain any information that names or uniquely identifies individual people or offensive content.

\begin{table}[!h]
\centering
\caption{\label{tab:datasets} Dataset statistics. }
\vspace{-0.2cm}
\resizebox{\columnwidth}{!}{%
\begin{tabular}{lccc}
\toprule
\textbf{Dataset} & \textbf{\# Users} & \textbf{\# Items} & \textbf{\# Rec history (train/val/test)} \\ \midrule
\textbf{Toys} & 4502 & 4943 & 80\% / 10\% / 10\% \\
\textbf{Beauty} & 22334 & 12094 & 80\% / 10\% / 10\% \\
\textbf{Sports} & 35393 & 18287 & 80\% / 10\% / 10\% \\
\bottomrule
\end{tabular}
}
\end{table}

\paragraph{Evaluation Metrics}
For sequential recommendation, we report the three most commonly-used evaluation metrics:  Mean Reciprocal Rank (MRR), Normalized Discounted Cumulative Gain (NDCG), Hits@5~\cite{jin2023lmindexer, rajput2023recommender}.  MRR averages the reciprocal rank of the first relevant item, emphasizing the importance of its position. NDCG evaluates ranking quality, weighting relevant items higher in the list using a logarithmic discount. Hits@5 measures the proportion of times a relevant item appears within the top 5 recommendations. These metrics offer complementary insights: MRR focuses on the first relevant item's rank, NDCG evaluates the overall ranking quality, and Hits@5 measures the presence of relevant items within the top recommendations. Together, they provide a comprehensive assessment of the recommendation system's performance.

\paragraph{Implementation Details} We use RQ-VAE as our indexer to generate discrete semantic IDs for each modality and T5 as our sequential recommender. For hyperparameter and computing resource details, please see \cref{sec:implementation}.

\paragraph{Baseline Methods} We compare our method with two categories of baselines: \textbf{(1)} Various sequential recommendation models, including popular approaches such as SASREC \cite{kang2018self} and generative recommendation methods with semantic IDs like LETTER \cite{jin2023lmindexer}, HCindexer \cite{wang2024learnable}, and TIGER(BERT) \cite{rajput2023recommender}. \textbf{(2)} To validate the efficacy of multimodal generative retrieval, we compare our methods against corresponding unimodal counterparts using TIGER as the main backbone. These baselines are denoted as TIGER(CLIP-text) and TIGER(CLIP-image), where the parentheses indicate the feature encoders for semantic IDs. Notably, our MGR-EF, MGR-LF, and \method also utilize TIGER as the main backbone, with CLIP-text and CLIP-image serving as text and image feature encoders, respectively.

\paragraph{Proposed Methods} We present MGR-EF and MGR-LF as two naive strategies for the MGR problem. Additionally, we report \method to demonstrate the results after incorporating our two key proposed designs: contrastive modality alignment and special tokens.

\begin{table*}[!htb]
\centering
\caption{\label{tab:results} \textbf{Performance of proposed \method and various baselines} across three sequential recommendation datasets. The best performance is highlighted in gray. \method achieves best performance across datasets.}
\resizebox{\linewidth}{!}{%
\begin{tabular}{lccc ccc ccc}
\toprule
\multicolumn{1}{c}{\multirow{2}{*}{\textbf{Method}}} & \multicolumn{3}{c}{\textbf{Toys}} & \multicolumn{3}{c}{\textbf{Beauty}} & \multicolumn{3}{c}{\textbf{Sports}} \\ \cmidrule(lr){2-4} \cmidrule(lr){5-7} \cmidrule(lr){8-10}
& \textbf{MRR} & \textbf{NDCG} & \textbf{Hits@5} & \textbf{MRR} & \textbf{NDCG} & \textbf{Hits@5} & \textbf{MRR} & \textbf{NDCG} & \textbf{Hits@5} \\ \midrule

LETTER~\cite{wang2024learnable} & 0.0169 & 0.0201 & 0.0300 & 0.0048 & 0.0059 & 0.0092 & 0.0018 & 0.0023 & 0.0036 \\

HC-indexer~\cite{jin2023lmindexer} & 0.0071 & 0.0088 & 0.0138 & 0.0074 & 0.0090 & 0.0138 & 0.0029 & 0.0036 & 0.0055 \\

SASRec ~\cite{kang2018self} & 0.0047 & 0.0056 & 0.0084 & 0.0084 & 0.0105 & 0.0167 & 0.0036 & 0.0042 & 0.0059 \\

TIGER(BERT) ~\cite{rajput2023recommender} & 0.0173 & 0.0212 & 0.0331 & 0.0077 & 0.0101 & 0.0173 & 0.0050 & 0.0062 & 0.0098 \\

\midrule

TIGER(CLIP-text) & 0.0217 & 0.0275 & 0.0451 & 0.0161 & 0.0199 & 0.0316 & \cellcolor{gray!30}0.0103 & 0.0127 & 0.0203 \\

TIGER(CLIP-image) & 0.0198 & 0.0245 & 0.0389 & 0.0143 & 0.0177 & 0.0283 & 0.0062 & 0.0077 & 0.0121 \\

\midrule

\textbf{MGR-EF (early fusion)} & 0.0231 & 0.0294 & 0.0486 & 0.0150 & 0.0187 & 0.0298 & 0.0086 & 0.0105 & 0.0163 \\

\textbf{MGR-LF (late fusion)} & 0.0220 & 0.0284 & 0.0482 & 0.0169 & 0.0210 & \cellcolor{gray!30}0.0336 & 0.0091 & 0.0114 & 0.0184 \\

\textbf{\method (proposed)} & \cellcolor{gray!30}0.0280 & \cellcolor{gray!30}0.0349 & \cellcolor{gray!30}0.0562 & \cellcolor{gray!30}0.0176 & \cellcolor{gray!30}0.0216 & 0.0334 & 0.0102 & \cellcolor{gray!30}0.0130 & \cellcolor{gray!30}0.0206 \\

\bottomrule
\end{tabular}
}
\end{table*}

\begin{figure}[t!]
\centering

  \includegraphics[width=0.98\columnwidth]{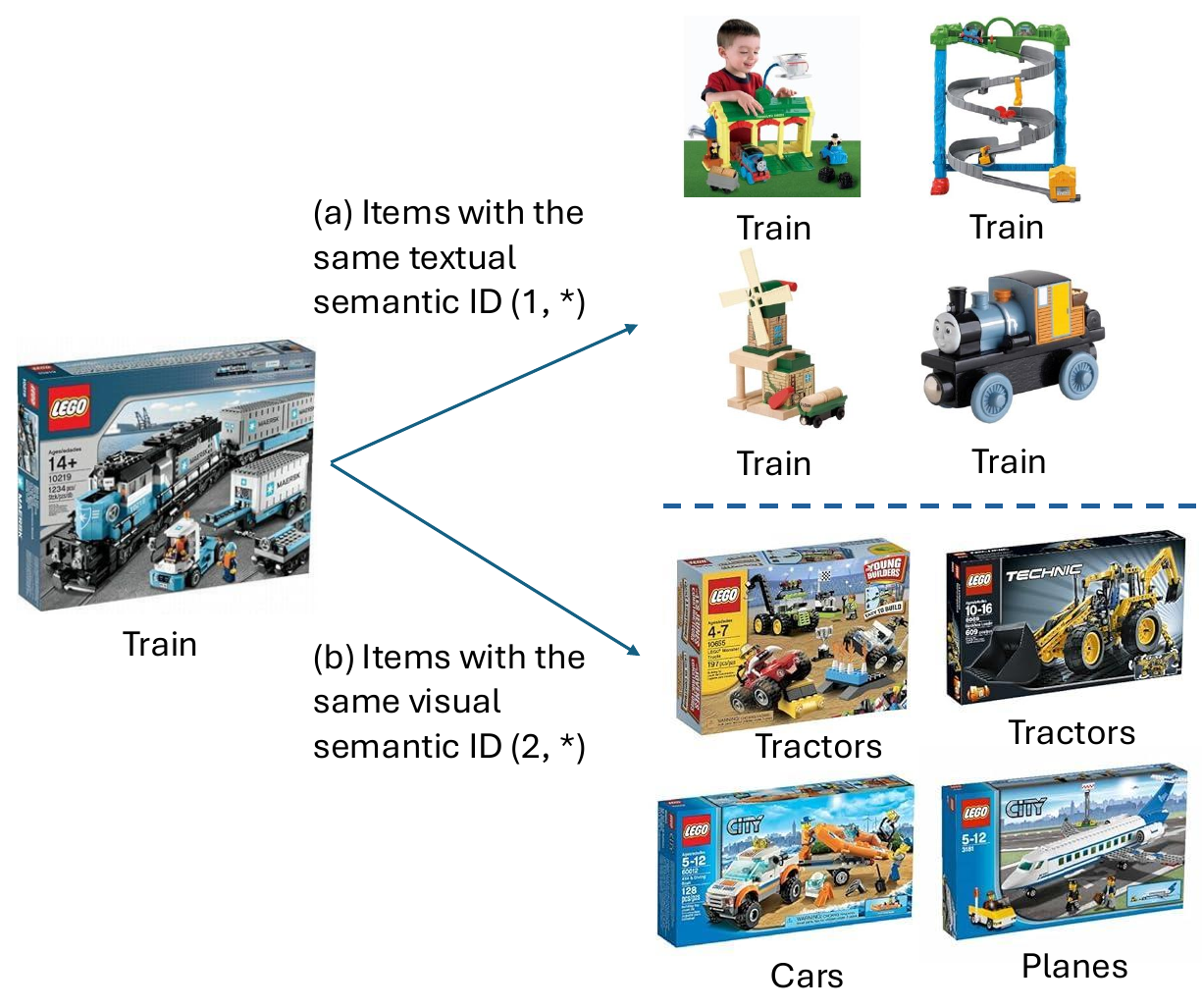}

\caption{{\textbf{Multimodality is needed: Semantic IDs of each modality captures different information.} Visualization of items with the same first semantic ID per modality wrt. the pivot example on Toys.} (a) focuses on capturing items with similar textual semantics with various visual lookings while (b) focuses on capturing items that look similar to each other visually. }

\label{fig:visual-example}
\end{figure}

\subsection{(RQ1) MGR vs. Unimodal GR}
\label{sec:overall_results}

\cref{tab:results} presents the overall experimental results. Even the most basic multimodal strategies, MGR-EF and MGR-LF, demonstrate substantial performance improvements over unimodal counterparts such as TIGER(CLIP-text) by 7\%. This underscores the value of extending existing unimodal generative retrieval frameworks to multimodal scenarios and highlights the significance of the MGR problem. Notably, LETTER exhibits poor performance in our experiments, attributable to the presence of unseen items without collaborative information in our datasets. As LETTER heavily relies on previously defined collaborative information, it fails to generate meaningful semantic IDs for these novel items.

Furthermore, \cref{fig:visual-example} provides qualitative examples of items grouped by shared first textual or visual semantic IDs. \cref{fig:visual-example}(a) demonstrates that textual semantic IDs capture items with similar textual meanings—in this case, all trains—while their visual representations vary significantly. Conversely, \cref{fig:visual-example}(b) illustrates that items sharing visual semantic IDs appear visually similar—all LEGO toy boxes—despite representing different types of toys such as tractors, cars, and planes. When recommending similar items, both textual and visual similarity should be considered, as some users may prefer similar trains, while others may be interested in visually similar LEGO toys representing different vehicles.

These findings demonstrate that leveraging semantic IDs from multiple modalities is essential for capturing diverse item relationships. This further underscores the importance of addressing the MGR problem to enhance generative recommendation in multimodal contexts.

\begin{figure*}[t!]
\centering
    \begin{subfigure}{0.3\linewidth}
    \centering
    \includegraphics[width=0.9\columnwidth]{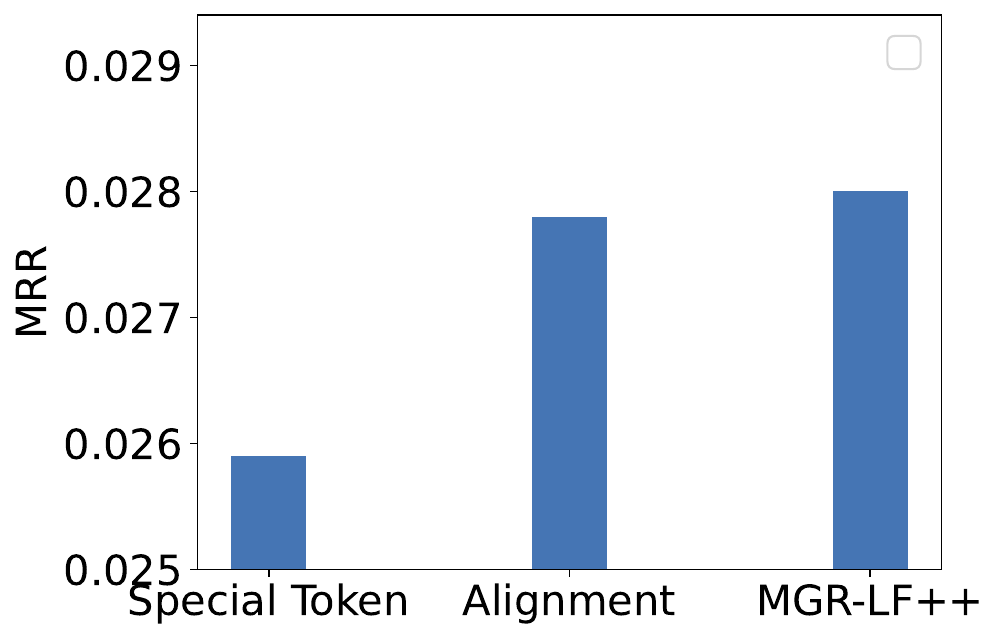}
    \caption{Toys}
    \end{subfigure}
    ~ 
    \begin{subfigure}{0.3\linewidth}
    \centering
    \includegraphics[width=0.9\columnwidth]{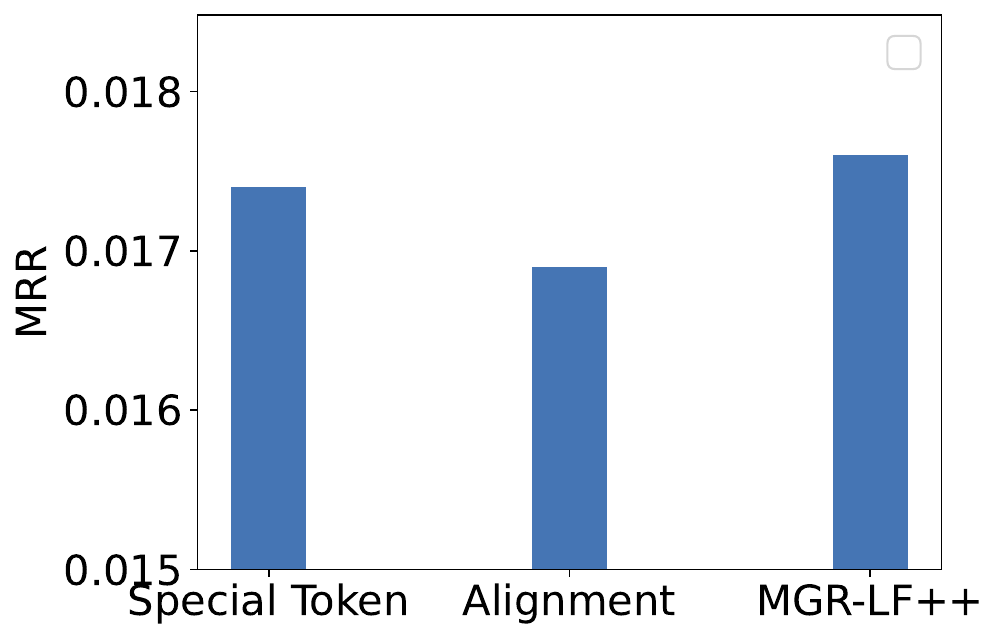}
    \caption{Beauty}
    \end{subfigure}
    ~
    \begin{subfigure}{0.3\linewidth}
    \centering~\includegraphics[width=0.9\columnwidth]{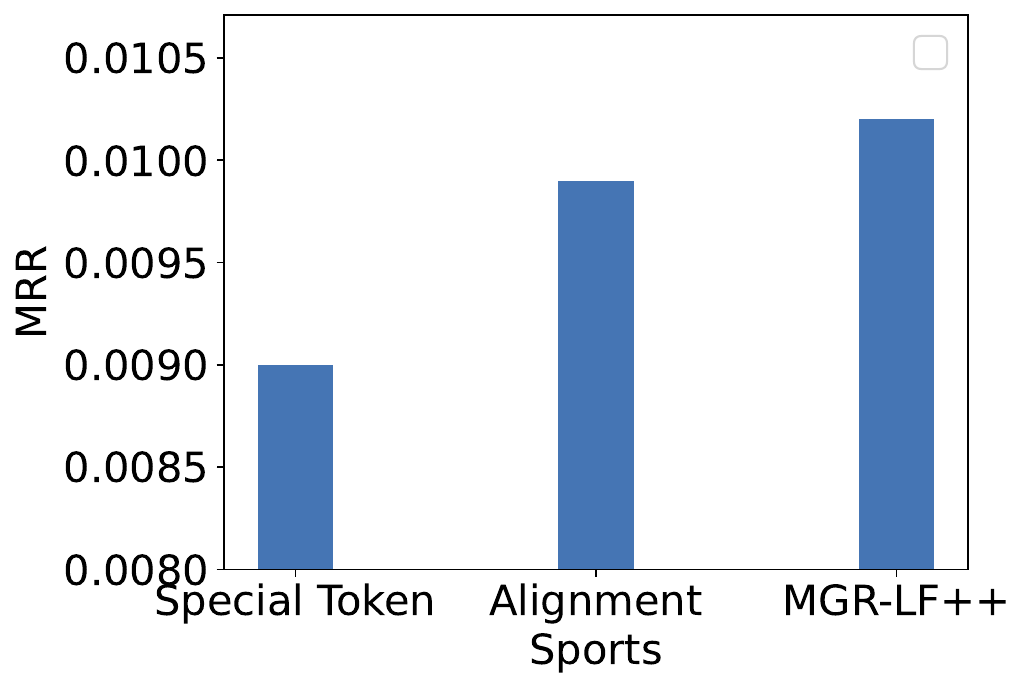}
    \caption{Sports}
    \end{subfigure}
\vspace{-0.05cm}
\caption{\textbf{Ablation Study on our two design choices across datasets.} Both of our proposed designs: special token
for cross-modal indication and contrastive modality are
important for the MGR problem.}

\label{fig:design-choices}
\end{figure*}

\begin{figure*}[t!]
\centering

\begin{subfigure}{0.45\linewidth}
\centering
\includegraphics[width=0.7\columnwidth]{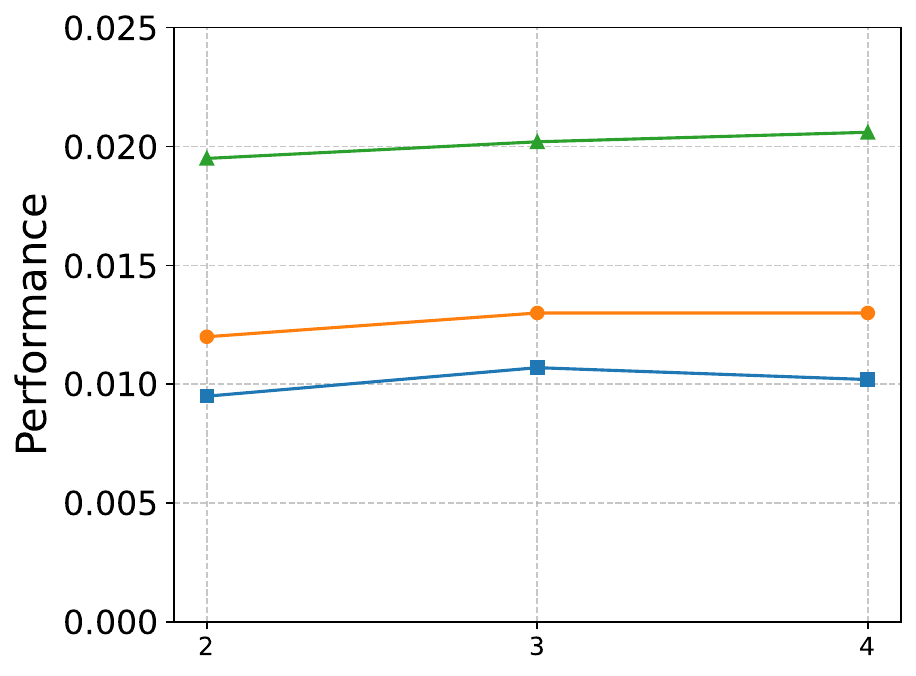}
\caption{Semantic ID Length Per Modality}
\end{subfigure}
~
\begin{subfigure}{0.45\linewidth}
\centering
\includegraphics[width=0.7\columnwidth]{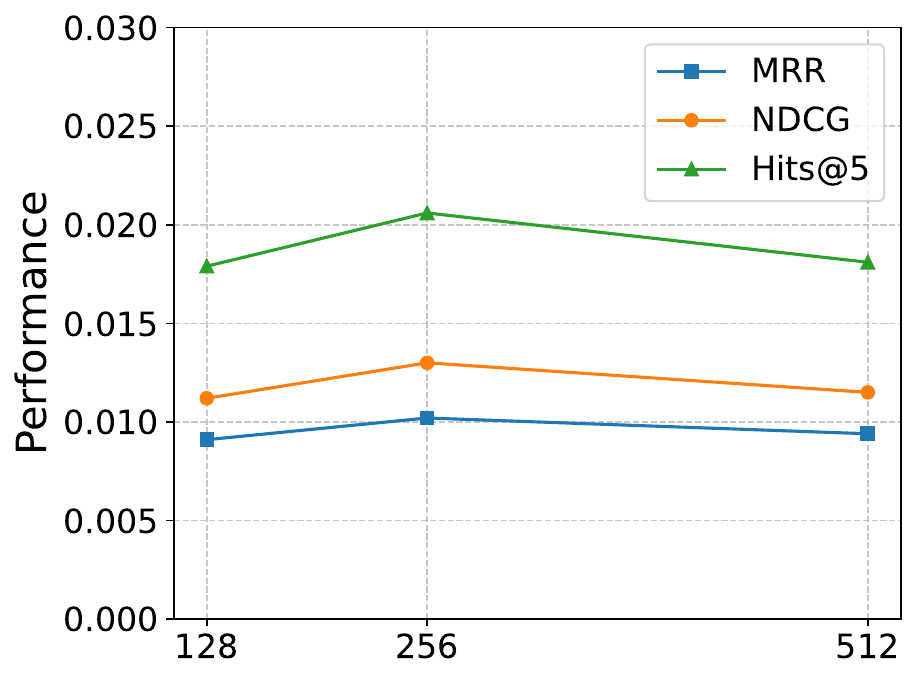}
\caption{Codebook Size}
\end{subfigure}

\vspace{-0.2cm}
\caption{
Influence of (a) semantic ID length and (b) codebook size on \method on Sports. For larger datasets, it is important to have a longer semantic ID length.
Setting codebook size to 256 gets the best performance.}

\label{fig:combined-ablation}
\vspace{-0.3cm}
\end{figure*}

\subsection{(RQ2) \method vs. Naive Strategies}

As evidenced in \cref{tab:results}, \method outperforms the basic multimodal strategies, MGR-EF and MGR-LF, across various datasets. \method achieves an average relative improvement of 20\% compared to MGR-LF. This significant performance boost demonstrates the effectiveness of our proposed designs: contrastive alignment pretraining and special tokens. These results suggest that \method effectively addresses the challenges of modality sensitivity and modality correspondences previously discussed.

\cref{fig:design-choices} provides a more detailed analysis of how each design choice affects performance, with "alignment" referring to contrastive modality alignment. Both special tokens and contrastive modality alignment contribute to substantial performance improvements, with their combination yielding the best overall performance. For Toys and Sports, contrastive modality alignment is more important while for Beauty, adding special tokens between modalities is more important.

\subsection{(RQ3) Ablation Studies}

We investigate the impact of semantic ID length and codebook size on our framework's performance. Using the same experimental setup as in Sections 4.1, we conduct ablation studies on the Sports dataset, our largest dataset. The results are visualized in \cref{fig:combined-ablation}.

\paragraph{Semantic ID Length} As illustrated in \cref{fig:combined-ablation}, longer semantic ID lengths generally yield better performance for larger datasets, despite slight variations across metrics. This improvement can be attributed to the richer representation provided by longer semantic IDs.

\paragraph{Codebook Size} Our analysis, also depicted in \cref{fig:combined-ablation}, reveals that a codebook size of 256 consistently produces the best overall performance across various metrics.

\section{Related Work}

\noindent \textbf{Sequential Recommendation} aims to predict a user's next interaction by modeling their historical behavior sequence. While traditional approaches relied on sequential pattern mining and Markov chains to capture interaction transitions ~\cite{fournier2017survey, yap2012effective, rendle2010factorizing}, recent advances adapt transformer-based models to model long-range dependencies in user interaction sequences ~\cite{kang2018self, zhu2024collaborative, sun2019bert4rec}. However, these methods predominantly focus on collaborative signals and sequential patterns, often overlooking item content. Generative retrieval, which captures item content via semantic IDs, has emerged as a new paradigm for sequential recommendation. In this work, inspired by the recent advances of language models, we use autoregressive language model as our sequential recommenders. 

\vspace{0.5em}
\noindent \textbf{Generative Recommendation},
which directly generates a ranked list of identifiers (IDs) for a given query using LLMs, has emerged as a new paradigm for information retrieval ~\cite{sun2024learning, rajput2023recommender, zheng2024adapting, deldjoo2024review, jin2023lmindexer}. At the core of these generative recommendation models is the construction of semantic IDs, a sequence of discrete ID numbers that captures the semantic meaning of a document. Current self-supervised semantic indexing methods generally follow a two-step process. In the first step, an off-the-shelf text encoder ~\cite{devlin-etal-2019-bert} encodes input documents and generates embedding representations for them. In the second step, an auto-encoder framework (e.g., RQ-VAE) is typically leveraged to generate semantic IDs for each document. For the auto-encoder framework, the encoder learns discrete latent variables for input documents and the decoder reconstructs input from these discrete variables ~\cite{rajput2023recommender, sun2024learning, jin2023lmindexer, wang2024learnable}. In this work, we extend the semantic ID generation pipeline to multimodal scenarios and uncover new challenges. Our work proposes new research questions and opens up new directions for generative recommendation.

\vspace{0.5em}
\noindent \textbf{Multimodal Representation Learning} aims to learn effective representations that can capture information across different modalities, such as text, images, and audio. Transformer-based models have shown remarkable success in multimodal feature learning ~\cite{xu2023multimodal, radford2021learning}. These models typically learn transferable visual representations by leveraging corresponding natural language supervision. Models like FLAVA ~\cite{singh2022flava} and Perceiver~\cite{jaegle2021perceiver} have demonstrated the effectiveness of jointly pre-training transformers on unpaired images and text, while CLIP has shown that contrastive objectives can effectively align representations from different modalities ~\cite{singh2022flava,jaegle2021perceiver,radford2021learning}.For more details, we refer readers to the survey~\cite{xu2023multimodal}.

\section{Conclusions}

Our work highlights the critical role of modality choice in generative recommendation (GR) systems, challenging the unimodal assumptions prevalent in prior research. By introducing and exploring the problem of Multimodal Generative Recommendation (MGR), we highlight the untapped potential of leveraging multimodal content to enhance generative recommendation performance in real-world scenarios. Our findings demonstrate that GR models can be sensitive to modality selection and highlights the key challenges associated with MGR, when naive early/late fusion strategy is used: modality sensitivity and modality correspondence. To address these, we propose MGR-LF++, an enhanced late fusion framework that significantly improves performance by incorporating contrastive modality alignment and special tokens to denote modality transitions. This research not only advances our understanding of GR in multimodal contexts but also lays a foundation for future work aimed at optimizing the use of diverse modalities in generative recommendation systems.

\section*{Limitations}

Our work showcases the promising potential of multimodal generative recommendation (MGR) and presents effective strategies, while also identifying several areas for further investigation.

First, our experiments reveal that early fusion is prone to modality sensitivity, where dominant modalities like text can overshadow others (e.g., images). Exploring a more balanced early fusion mechanism that can handle this challenge is worth investigating further.

Second, the datasets used in our experiments may not fully represent the diversity and complexity of real-world multimodal recommendation scenarios. For example, visual data in some datasets may be more informative than textual data or vice versa, which could bias the results toward certain modalities. Moreover, we primarily focused on text and image modalities due to the lack of publicly available datasets that have additional modalities (e.g., audio, video, tabular data). Future work should introduce and explore more diverse datasets to validate the generalizability of our findings.

\bibliographystyle{acl_natbib}
\bibliography{anthology, custom}

\appendix
\section{Appendix}

\subsection{AMI score}
\label{sec:ami}

The Adjusted Mutual Information (AMI) score ~\cite{vinh2009information} is a measure used in statistics and information theory to quantify the agreement between two clusters (in our
experiments, the two clusters refer to ground truth category clusters and Semantic ID clusters) while correcting for chance. It is an adjustment of the Mutual Information (MI) score that accounts for the fact that MI is generally higher for clusters with a larger number of clusters, thus providing a normalized score that is more comparable across different clusters. Here we use AMI to calculate how much overlap the semantic IDs are across modalities. And since semantic IDs generated by RQ-VAE are hierarchical, we select the first semantic ID for each modality to compute the AMI score.

\subsection{Implementation Details}
\label{sec:implementation}
For generative language model that are used as sequential recommender, we use T5-small and T5-base.
When generating semantic IDs, the textual information
(title \& description) is encoded via the text encoder of CLIP and the visual information is encoder via the image encoder of CLIP to generate semantic embeddings for each modality. Take the embeddings as inputs, an rq-VAE indexer is used to generate semantic IDs for each item.  The
baselines use the same T5 checkpoint for fair comparison. Given the size of the datasets are different, we train 10,000 steps for toys, 15,000 steps for beauty and 20,000 steps for sports. For contrastive alignment pretraining, we first pretrain the corresponding steps using the contrastive objective and then use the pretrained checkpoint and train it directly on sequential recommedendation. We search learning rate in  {1e-2, 1e-3,
1e-4}. The maximum input text length is set to 1024 and all experiments are run on an 4
A100 40G machine. The number of beams for beam search
is set to 20. Due to computation costs, all experiments are run once.

\end{document}